\documentclass[english,aps]{revtex4}
\usepackage[T1]{fontenc}
\usepackage[latin9]{inputenc}
\usepackage[a4paper]{geometry}
\usepackage{amsmath}
\usepackage{graphicx}
\usepackage{esint}

\makeatletter
\@ifundefined{textcolor}{}
{%
 \definecolor{BLACK}{gray}{0}
 \definecolor{WHITE}{gray}{1}
 \definecolor{RED}{rgb}{1,0,0}
 \definecolor{GREEN}{rgb}{0,1,0}
 \definecolor{BLUE}{rgb}{0,0,1}
 \definecolor{CYAN}{cmyk}{1,0,0,0}
 \definecolor{MAGENTA}{cmyk}{0,1,0,0}
 \definecolor{YELLOW}{cmyk}{0,0,1,0}
}

\makeatother

\usepackage{babel}
\begin{document}

\title{Electromagnetic fields with electric and chiral magnetic conductivities
in heavy ion collisions}

\author{Hui Li}

\affiliation{Interdisciplinary Center for Theoretical Study and Department of
Modern Physics, University of Science and Technology of China, Hefei,
Anhui 230026, China}

\author{Xin-li Sheng}

\affiliation{Interdisciplinary Center for Theoretical Study and Department of
Modern Physics, University of Science and Technology of China, Hefei,
Anhui 230026, China}

\author{Qun Wang}

\affiliation{Interdisciplinary Center for Theoretical Study and Department of
Modern Physics, University of Science and Technology of China, Hefei,
Anhui 230026, China}
\begin{abstract}
We derive analytic formula for electric and magnetic fields produced
by a moving charged particle in a conducting medium with the electric
conductivity $\sigma$ and the chiral magnetic conductivity $\sigma_{\chi}$.
We use the Green function method and assume that $\sigma_{\chi}$
is much smaller than $\sigma$. The compact algebraic expressions
for electric and magnetic fields without any integrals are obtained.
They recover the Lienard-Wiechert formula at vanishing conductivities.
Exact numerical solutions are also found for any values of $\sigma$
and{\normalsize{} $\sigma_{\chi}$} and are compared to analytic results.
Both numerical and analytic results agree very well for the scale
of high energy heavy ion collisions. The space-time profiles of electromagnetic
fields in non-central Au+Au collisions have been calculated based
on these analytic formula as well as exact numerical solutions. 
\end{abstract}
\maketitle

\section{Introduction}

Strong electromagnetic fields are generated in peripheral heavy-ion
collisions (HIC), which provides a good opportunity for studying rich
phenomena related to strong fields. At the collisional energy $\sqrt{s}$
per nucleon which is much larger than the nucleon mass $m_{n}$, the
nucleons are moving with the velocity $v=\sqrt{(s-m_{n}^{2})/s}\sim1-m_{n}^{2}/(2s)$
which is almost speed of light with a large Lorentz contraction factor
$\gamma=1/\sqrt{1-v^{2}/c^{2}}\sim\sqrt{s}/m_{n}$. The typical electric
field in the co-moving frame of one nucleus can be estimated by the
Coulomb law, $Ze/R_{A}^{2}$, with $Z$ and $R_{A}$ being the proton
number and the radius of the nucleus respectively. The magnetic field
in the lab frame can be approximated as the product of the Lorentz
factor and the electric field in the co-moving frame of the nucleus,
$eB\sim\gamma vZe^{2}/R_{A}^{2}$. In Au+Au collisions at the Relativistic
Heavy Ion Collider (RHIC) at $\sqrt{s}=200$ GeV, the peak value of
the magnetic field at the moment of the collision is about $5m_{\pi}^{2}$
($m_{\pi}$: pion mass) or $1.4\times10^{18}$ Gauss. In Pb+Pb collisions
at the Large Hadron Collider (LHC) at $\sqrt{s}=2.76$ TeV, the peak
value of the magnetic field can be 10 times as large as at RHIC. 

Since the magnitude of the electromagnetic fields enter the regime
of strong interaction, the effects of such enormous fields are expected
to be observable in the final hadronic events in HIC. In recent years
there have been many efforts to investigate such effects, among which
the interplay between strong magnetic fields and quantum anomaly leads
to a group of related phenomena, such as the Chiral Magnetic Effect
(CME) \cite{Kharzeev:2007jp,Fukushima:2008xe}, the Chiral Vortical
Effect \cite{Son:2009tf,Kharzeev:2010gr}, and the Chiral Magnetic
Wave \cite{Burnier:2011bf}, the Chiral Vortical Wave \cite{Jiang:2015cva},
etc.. For reviews of recent developments, see, e.g. \cite{Kharzeev:2013jha,Kharzeev:2015znc}.
All these effects are related to chiral properties of fermions, especially
massless fermions or chiral fermions. The movement of chiral fermions
can be described by the chiral kinetic equations which incorporate
structures of Berry phase and monopole in momentum space \cite{Son:2012wh,Stephanov:2012ki,Gao:2012ix,Chen:2012ca,Son:2012zy,Chen:2013iga,Manuel:2013zaa,Duval:2014ppa,Chen:2014cla,Satow:2014lva}.
The charge separation effect observed in the STAR and ALICE experiments
can be well described by the CME \cite{Abelev:2009ac,Abelev:2009ad,Abelev:2012pa},
but no definite conclusion has been made that the charge separation
effect results unambiguously and exclusively from the CME instead
of the collective expansion of the fireball. 

As a starting point to study these phenomena, one must know the space-time
profile of electromagnetic fields in HIC. Several earlier calculations
\cite{Kharzeev:2007jp,Skokov:2009qp,Voronyuk:2011jd,Deng:2012pc}
as well as later calculations including event-by-event fluctuations
\cite{Bloczynski:2012en,Roy:2015coa} show that the electromagnetic
fields peak almost at the time of collision and disappear in very
short time after the collision. For example, the magnetic field along
the global angular momentum falls rapidily by $\sim1/t^{3}$. At $\sqrt{s}=200$
GeV, it drops by two to three orders of magnitude in about 0.5 fm/c
from the collision time. If this is the case, one cannot expect a
sizable influence on the final state hadrons in late time from such
short pulses of magmetic fields. However the medium effects have not
be considered in these calculations. The main response of the plasma
to the fields is the electric conduction. The electric conductivity
is proportional to plasma temperature, which is a function of time
because the plasma is expanding. In strong coupling regime, electric
conductivity can be calculated by lattice gauge theory \cite{Ding:2010ga,Aarts:2014nba}
and holographic models \cite{Finazzo:2013efa}. Ohm's currents will
be induced in the plasma and slow down the decrease of the fields
\cite{Tuchin:2013apa,McLerran:2013hla}. To study the CME effect,
one has to include the CME conductivity $\sigma_{\chi}$. The electromagnetic
fields produced by a point charge with $\sigma_{\chi}$ and $\sigma$
have been calculated analytically in Ref. \cite{Tuchin:2014iua} but
only for the relativitic limit ($v=1$). The numerical and analytic
calculations with $\sigma$ but without $\sigma_{\chi}$ were done
in Ref. \cite{Tuchin:2013apa,McLerran:2013hla}. The directed flow
of charged hadrons in HIC has been studied with non-vanishing $\sigma$
but without $\sigma_{\chi}$ in Ref. \cite{Gursoy:2014aka} by calculating
the velocity shift of each fluid cell due to electromagnetic force
in the hydrodynamic evolution. 

In this paper, we will solve the Maxwell equations both analytically
and numerically for a moving point charge in a conducting medium with
non-vanishing $\sigma$ and $\sigma_{\chi}$. We use the method of
the Green functions under the condition $\sigma_{\chi}\ll\sigma$
which is valid for high energy HIC. Analytic expressions of electric
and magnetic fields are given for finite $\sigma$ and small values
of $\sigma_{\chi}$ without taking the relativistic limit ($v=1$).
The numerical results for finite $\sigma$ and $\sigma_{\chi}$ agree
perfectly with the analytic results. Finally we carry out the numerical
calculations for the electromagnetic fields in non-central Au+Au collisions
at $\sqrt{s}=200$ GeV. Normally one uses the AMPT model \cite{Lin:2004en},
the HIJING model \cite{Wang:1991hta}, or the UrQMD model \cite{Bass:1998ca}
to simulate the collision processes and to calculate the electromagnetic
fields. In this paper, we will use the UrQMD model to give the space-time
and momentum configuration of charged particles in HIC the calculations
with vanishing $\sigma$ and $\sigma_{\chi}$, but we will use a kinematic
model for participant nucleons with non-vanishing $\sigma$ and $\sigma_{\chi}$.
Generally the strong magnetic fields will influence the evolution
of the particle system \cite{Pu:2016ayh}, which we will not consider
in our calculations.  

The paper is organized as follows. In Section \ref{sec:formal}, we
give the formal solution to the Maxwell equations with $\sigma$ and
$\sigma_{\chi}$ using the method of Green functions. In Section \ref{sec:mag}
and \ref{sec:elect}, we derive analytic expressions for magnetic
and electric fields of a point charge respectively. We give in Section
\ref{sec:numerical} numerical results for electromagnetic fields
produced in non-central Au+Au collisions at $\sqrt{s}=200$ GeV. A
summary of results is given in Section \ref{sec:summary}. 

We will adopt following conventions for three-dimensional (3D) or
two-dimensional (2D) vectors. We will use Roman letters in boldface
for 3D or 2D vectors. In Cartesian coordinates, three orthogonal components
of a 3D vector are denoted as plain Roman letters with subscripts
$x,y,z$. A point in coordinate space is written as $\mathbf{x}=(x,y,z)=(\mathbf{x}_{T},z)$,
where $\mathbf{x}_{T}$ represents its 2D component. Similarly a momentum
is written as $\mathbf{k}=(k_{x},k_{y},k_{z})=(\mathbf{k}_{T},k_{z})$.
The vectors of the electric and magnetic fields are written as $\mathbf{E}=(E_{x},E_{y},E_{z})$
and $\mathbf{B}=(B_{x},B_{y},B_{z})$. We will also use cylindrical
coordiante whose longitudinal component is chosen to be the third
component of Cartesian coordinate, e.g., $\mathbf{x}=(x_{T},\phi,z)$
with $x_{T}=|\mathbf{x}_{T}|$.

\section{Field Equations and their formal solutions}

\label{sec:formal}We consider an infinite homogeneous medium whose
conducting property can be described by a constant electric conductivity
$\sigma$ and a constant chiral magnetic conductivity $\sigma_{\chi}$.
These requirements give us the most simplified model for conducting
medium with the chiral magnetic effect (CME). In this medium, the
total currents can be decomposed into three parts, the external current,
the Ohm's current ($\sigma\mathbf{E}$) induced by the electric field
$\mathbf{E}$, and the chiral magnetic current ($\sigma_{\chi}\mathbf{B}$)
induced by the magnetic field $\mathbf{B}$. The Maxwell equations
read
\begin{eqnarray}
\nabla\cdot\mathbf{E} & = & \frac{\rho_{\mathrm{ext}}}{\epsilon},\nonumber \\
\nabla\cdot\mathbf{B} & = & 0,\nonumber \\
\nabla\times\mathbf{E} & = & -\partial_{t}\mathbf{B},\nonumber \\
\nabla\times\mathbf{B} & = & \partial_{t}\mathbf{E}+\mathbf{J}_{\mathrm{ext}}+\sigma\mathbf{E}+\sigma_{\chi}\mathbf{B},\label{eq:field-eq-sigma}
\end{eqnarray}
where $\rho_{\mathrm{ext}}$ and $\mathbf{J}_{\mathrm{ext}}$ denote
the external charge and current densities respectively. One should
note that in general the permittivity, $\epsilon(\omega)=1+i\sigma/\omega$,
depends on the frequency. Taking curl of the third and the fourth
line in Eq. (\ref{eq:field-eq-sigma}) and using the first and the
second line, we obtain,
\begin{eqnarray}
\left(\nabla^{2}-\partial_{t}^{2}-\sigma\partial_{t}\right)\mathbf{B}+\sigma_{\chi}\nabla\times\mathbf{B} & = & -\nabla\times\mathbf{J}_{\mathrm{ext}},\nonumber \\
\left(\nabla^{2}-\partial_{t}^{2}-\sigma\partial_{t}\right)\mathbf{E}+\sigma_{\chi}\nabla\times\mathbf{E} & = & \frac{1}{\epsilon}\nabla\rho_{\mathrm{ext}}+\partial_{t}\mathbf{J}_{\mathrm{ext}}.\label{eq:eom-sigma}
\end{eqnarray}
It is obvious that both the magnetic and electric fields satisfy the
same system of partial differential equations,
\begin{equation}
\hat{L}\mathbf{F}\left(t,\mathbf{x}\right)+\sigma_{\chi}\nabla\times\mathbf{F}\left(t,\mathbf{x}\right)=\mathbf{f}\left(t,\mathbf{x}\right).\label{eq:eom-1}
\end{equation}
Here $\mathbf{F}(t,\mathbf{x})$ is a vector representing $\mathbf{B}$
or $\mathbf{E}$. The partial differential operator is defined as
$\hat{L}=\nabla^{2}-\partial_{t}^{2}-\sigma\partial_{t}$. The function
$\mathbf{f}(t,\mathbf{x})$ on the right-hand-side stands for the
source terms in Eq. (\ref{eq:eom-sigma}). We can also write Eq. (\ref{eq:eom-1})
in a matrix form in terms of three components of $\mathbf{F}=(F_{x},F_{y},F_{z})$
and $\mathbf{f}=(f_{x},f_{y},f_{z})$, 
\begin{equation}
\left(\begin{array}{ccc}
\hat{L} & -\sigma_{\chi}\partial_{z} & \sigma_{\chi}\partial_{y}\\
\sigma_{\chi}\partial_{z} & \hat{L} & -\sigma_{\chi}\partial_{x}\\
-\sigma_{\chi}\partial_{y} & \sigma_{\chi}\partial_{x} & \hat{L}
\end{array}\right)\left(\begin{array}{c}
F_{x}\\
F_{y}\\
F_{z}
\end{array}\right)\left(t,\mathbf{x}\right)=\left(\begin{array}{c}
f_{x}\\
f_{y}\\
f_{z}
\end{array}\right)\left(t,\mathbf{x}\right).\label{eq:matrix-eq}
\end{equation}
where we have used the shorthand notation $\nabla=(\partial/\partial x,\partial/\partial y,\partial/\partial z)\equiv(\partial_{x},\partial_{y},\partial_{z})$. 

Now we are at the point to solve the above equation. To this end,
it is convenient to work in momentum space and expand $\mathbf{F}\left(t,\mathbf{x}\right)$
and $\mathbf{f}\left(t,\mathbf{x}\right)$ as 
\begin{eqnarray}
\mathbf{F}\left(t,\mathbf{x}\right) & = & \int\frac{d\omega d^{3}\mathbf{k}}{(2\pi)^{4}}\mathrm{e}^{-i\omega t+i\mathbf{k}\cdot\mathbf{x}}\mathbf{F}(\omega,\mathbf{k}),\nonumber \\
\mathbf{f}\left(t,\mathbf{x}\right) & = & \int\frac{d\omega d^{3}\mathbf{k}}{(2\pi)^{4}}\mathrm{e}^{-i\omega t+i\mathbf{k}\cdot\mathbf{x}}\mathbf{f}(\omega,\mathbf{k}).\label{eq:fourier}
\end{eqnarray}
Insert the above expressions into Eq. (\ref{eq:matrix-eq}), we obtain
by making replacement $\partial_{t}\rightarrow-i\omega$, $\nabla\rightarrow i\mathbf{k}$,
\begin{equation}
\left(\begin{array}{ccc}
L & -i\sigma_{\chi}k_{z} & i\sigma_{\chi}k_{y}\\
i\sigma_{\chi}k_{z} & L & -i\sigma_{\chi}k_{x}\\
-i\sigma_{\chi}k_{y} & i\sigma_{\chi}k_{x} & L
\end{array}\right)\left(\begin{array}{c}
F_{x}\\
F_{y}\\
F_{z}
\end{array}\right)(\omega,\mathbf{k})=\left(\begin{array}{c}
f_{x}\\
f_{y}\\
f_{z}
\end{array}\right)(\omega,\mathbf{k}),\label{eq:matrix-eq-mom}
\end{equation}
where $L=\omega^{2}+i\sigma\omega-k^{2}$ and $k=|\mathbf{k}|$. We
can write the coefficient matrix in a compact form,
\begin{equation}
M_{ij}=L\delta_{ij}-i\sigma_{\chi}\epsilon_{ijl}k_{l},\label{eq:compact-M}
\end{equation}
with the determinant 
\begin{equation}
\det M=L(L^{2}-\sigma_{\chi}^{2}k^{2}).
\end{equation}
In Eq. (\ref{eq:compact-M}) we have used the notation $\mathbf{k}=(k_{x},k_{y},k_{z})=(k_{1},k_{2},k_{3})$.
If $\det M\neq0$, we can get the inverse of $M$ given by its adjoint
matrix divided by its determinant,
\begin{equation}
M^{-1}=\frac{1}{\mathrm{det}M}\left(\begin{array}{ccc}
L^{2}-\sigma_{\chi}^{2}k_{x}^{2} & iL\sigma_{\chi}k_{z}-\sigma_{\chi}^{2}k_{x}k_{y} & -iL\sigma_{\chi}k_{y}-\sigma_{\chi}^{2}k_{x}k_{z}\\
-iL\sigma_{\chi}k_{z}-\sigma_{\chi}^{2}k_{x}k_{y} & L^{2}-\sigma_{\chi}^{2}k_{y}^{2} & iL\sigma_{\chi}k_{x}-\sigma_{\chi}^{2}k_{y}k_{z}\\
iL\sigma_{\chi}k_{y}-\sigma_{\chi}^{2}k_{x}k_{z} & -iL\sigma_{\chi}k_{x}-\sigma_{\chi}^{2}k_{y}k_{z} & L^{2}-\sigma_{\chi}^{2}k_{z}^{2}
\end{array}\right).
\end{equation}
With $M^{-1}$ we can write down the solution to Eq. (\ref{eq:matrix-eq-mom})
as 
\begin{equation}
\mathbf{F}\left(\omega,\mathbf{k}\right)=\frac{1}{L^{2}-\sigma_{\chi}^{2}k^{2}}\left[L\mathbf{f}(\omega,\mathbf{k})-i\sigma_{\chi}\mathbf{k}\times\mathbf{f}(\omega,\mathbf{k})\right]-\frac{\sigma_{\chi}^{2}}{L(L^{2}-\sigma_{\chi}^{2}k^{2})}\mathbf{k}[\mathbf{k}\cdot\mathbf{f}(\omega,\mathbf{k})],\label{eq:solution}
\end{equation}
where the source terms $\mathbf{f}\left(\omega,\mathbf{k}\right)$
are given by 
\begin{equation}
\mathbf{f}(\omega,\mathbf{k})=\begin{cases}
-i\mathbf{k}\times\mathbf{\mathbf{J}}_{\mathrm{ext}}(\omega,\mathbf{k}), & \mathrm{for}\:\mathbf{B}\\
i\mathbf{k}\frac{\rho_{\mathrm{ext}}(\omega,\mathbf{k})}{1+i\sigma/\omega}-i\omega\mathbf{\mathbf{J}}_{\mathrm{ext}}(\omega,\mathbf{k}), & \mathrm{for}\:\mathbf{E}
\end{cases}\label{eq:source}
\end{equation}
We note that the second term $\sim\sigma_{\chi}^{2}\mathbf{k}[\mathbf{k}\cdot\mathbf{f}(\omega,\mathbf{k})]$
in Eq. (\ref{eq:solution}) is obviously vanishing for $\mathbf{B}$,
but it is not vanishing for $\mathbf{E}$. Using the charge conservation
equation $\partial\rho_{\mathrm{ext}}/\partial t+\nabla\cdot\mathbf{J}_{\mathrm{ext}}=0$,
this term for $\mathbf{E}$ is proportional to $\sim[(L^{2}-\sigma_{\chi}^{2}k^{2})(1+i\sigma/\omega)]^{-1}$.
From the poles of $\mathbf{F}\left(\omega,\mathbf{k}\right)$, we
can obtain the dispersion relations $\omega(\mathbf{k})$ for collective
modes of electromagnetic fields. The poles of the first term (or of
$\mathbf{B}$) in Eq. (\ref{eq:solution}) are given by the roots
of $L^{2}-\sigma_{\chi}^{2}k^{2}=0$, which are $\omega_{s_{1}s_{2}}=-i\sigma/2+s_{1}\sqrt{k^{2}+s_{2}k\sigma_{\chi}-\sigma^{2}/4}$
($s_{1},s_{2}=\pm1$). For $\mathbf{E}$, the second term in Eq. (\ref{eq:solution})
introduces an additional pole $\omega=-i\sigma$ besides $\omega_{s_{1}s_{2}}$.
These poles give the collective modes of the fields without external
sources, where $\omega$ and $\mathbf{k}$ are independent variables.
For external charges with $\rho_{\mathrm{ext}}$ and $\mathbf{\mathbf{J}}_{\mathrm{ext}}$,
the dispersion relations will be modified due to additional relations
between $\omega$ and $\mathbf{k}$, e.g., in the next section we
will consider a point charge moving along the z-direction which introduces
the constraint $\omega=vk_{z}$.

\section{Magnetic Fields of a moving charge\label{sec:mag}}

\subsection{Integration over the polar angle and longitudinal momentum}

\textbf{\Large \label{2.-Magnetic-Fields}}In this section, we will
derive an analytical expression for the magnetic field of a charged
particle. Without loss of generality, we consider the situation that
the charged particle (with charge $Q$) moves along the third axis
direction. More general cases along arbitrary directions can be obtained
by rotation. In heavy ion collisions, generally the CME conductivity
is a small quantity compared to the electric one. The charge density
and the current density read,
\begin{eqnarray}
\rho\left(t,\mathbf{x}\right) & = & Q\delta(x)\delta(y)\delta(z-vt),\nonumber \\
\mathbf{J}\left(t,\mathbf{x}\right) & = & Qv\delta(x)\delta(y)\delta(z-vt)\mathbf{e}_{z}.\label{eq:source-x-space}
\end{eqnarray}
In momentum space, they are in the form 
\begin{eqnarray}
\rho(\omega,\mathbf{k}) & = & 2\pi Q\delta(\omega-k_{z}v),\nonumber \\
\mathbf{J}(\omega,\mathbf{k}) & = & 2\pi Qv\delta(\omega-k_{z}v)\mathbf{e}_{z}.\label{eq:current density}
\end{eqnarray}
Here we denote three directions in flat coordinate space as $(\mathbf{e}_{x},\mathbf{e}_{y},\mathbf{e}_{z})$.
In cylindrical coordinates, we denote three orthorgonal directions
as $(\mathbf{e}_{r},\mathbf{e}_{\phi},\mathbf{e}_{z})$. Inserting
Eq. (\ref{eq:current density}) into Eqs. (\ref{eq:solution},\ref{eq:source}),
we obtain the magnetic field in momentum space,
\begin{eqnarray}
\left(\begin{array}{c}
B_{x}\\
B_{y}\\
B_{z}
\end{array}\right)(\omega,\mathbf{k}) & = & -2\pi iQv\frac{\delta(\omega-k_{z}v)}{L^{2}-\sigma_{\chi}^{2}k^{2}}\left(\begin{array}{c}
Lk_{y}-i\sigma_{\chi}k_{x}k_{z}\\
-Lk_{x}-i\sigma_{\chi}k_{y}k_{z}\\
i\sigma_{\chi}(k_{x}^{2}+k_{y}^{2})
\end{array}\right).\label{eq:b-mom}
\end{eqnarray}

We can transform Eq. (\ref{eq:b-mom}) back to coordinate space. This
involves integration over $\omega$ and $\mathbf{k}$. Since the charged
particle moves along the third direction, it is convenient to work
in the cylindrical coordinate $(r,\phi,z)$. So we can write $\mathbf{k}\cdot\mathbf{x}=\mathbf{k}_{T}\cdot\mathbf{x}_{T}+k_{z}z=k_{T}x_{T}\cos\theta+k_{z}z$,
where we have assumed that the angle between $\mathbf{x}_{T}$ and
$\mathbf{k}_{T}$ is $\theta$. We can easily integrate over $k_{z}$
from Eq. (\ref{eq:b-mom}), which removes the delta function with
$k_{z}$ being set to $\omega/v$ in the integrand, 
\begin{eqnarray}
\left(\begin{array}{c}
B_{r}\\
B_{\phi}\\
B_{z}
\end{array}\right)(t,\mathbf{x}) & = & -iQ\int\frac{d\omega d\theta dk_{T}}{(2\pi)^{3}}\mathrm{e}^{-i\omega(t-z/v)+ik_{T}x_{T}\mathrm{cos}\theta}\frac{k_{T}^{2}}{D(\omega,k_{T})}\nonumber \\
 &  & \times\left[L(\omega,k_{T})\left(\begin{array}{c}
\mathrm{sin}\theta\\
-\mathrm{cos}\theta\\
0
\end{array}\right)+\frac{i\sigma_{\chi}}{v}\left(\begin{array}{c}
-\omega\mathrm{cos}\theta\\
-\omega\mathrm{sin}\theta\\
vk_{T}
\end{array}\right)\right],\label{eq:B-cylind}
\end{eqnarray}
where we have chosen that $\mathbf{x}_{T}$ is along $\mathbf{e}_{x}$,
so $\mathbf{e}_{r}$ ($\mathbf{e}_{x}$) is in the direction of $\mathbf{x}_{T}$
and $\mathbf{e}_{\phi}$ ($\mathbf{e}_{y}$) is in the direction of
$\mathbf{e}_{z}\times\mathbf{e}_{r}$. We have used in Eq. (\ref{eq:B-cylind})
\begin{eqnarray}
L(\omega,k_{T}) & = & -\frac{1}{v^{2}\gamma^{2}}\omega^{2}+i\sigma\omega-k_{T}^{2},\nonumber \\
D(\omega,k_{T}) & = & L^{2}(\omega,k_{T})-\sigma_{\chi}^{2}\frac{\omega^{2}}{v^{2}}-\sigma_{\chi}^{2}k_{T}^{2}.\label{eq:denominator}
\end{eqnarray}
Integration over $\theta$ can be done using cylindrical Bessel functions,
$\int_{0}^{2\pi}d\theta\mathrm{e}^{ik_{T}x_{T}\cos\theta}=2\pi J_{0}(k_{T}x_{T})$
and $\int_{0}^{2\pi}d\theta\mathrm{e}^{ik_{T}x_{T}\cos\theta}\cos\theta=2\pi iJ_{1}(k_{T}x_{T})$.
Inserting these into Eq. (\ref{eq:B-cylind}), we can have a simple
form, 
\begin{eqnarray}
\mathbf{B}(t,\mathbf{x}) & = & -Q\int\frac{d\omega dk_{T}}{(2\pi)^{2}}\:\frac{\mathbf{B}^{\prime}(\omega,k_{T})}{D(\omega,k_{T})},\label{eq:int-omega-k}
\end{eqnarray}
where $\mathbf{B}^{\prime}(\omega,k_{T})$ can be defined in cylindrical
coordiates as follows 
\begin{eqnarray}
\left(\begin{array}{c}
B_{r}^{\prime}\\
B_{\phi}^{\prime}\\
B_{z}^{\prime}
\end{array}\right)(\omega,k_{T}) & \equiv & k_{T}^{2}\mathrm{e}^{-i\omega(t-z/v)}\nonumber \\
 &  & \times\left[L(\omega,k_{T})\left(\begin{array}{c}
0\\
J_{1}(k_{T}x_{T})\\
0
\end{array}\right)+\frac{\sigma_{\chi}}{v}\left(\begin{array}{c}
i\omega J_{1}(k_{T}x_{T})\\
0\\
-vk_{T}J_{0}(k_{T}x_{T})
\end{array}\right)\right],\label{eq:numerator}
\end{eqnarray}
where $J_{0}$ and $J_{1}$ are Bessel functions of the first kind.
Note that the integral over $\omega$ in Eq. (\ref{eq:int-omega-k})
is from $-\infty$ to $+\infty$. We can easily prove that the right-hand
side of Eq. (\ref{eq:int-omega-k}) is a real number. We see in the
integrand that $\omega$ is always accompanied by an imaginary unit
$i$. If we replace $\omega$ with $-\omega$ in the integrand, we
will get exactly its complex conjugate, so the the integral over $\omega$
in Eq. (\ref{eq:int-omega-k}) can be replaced by an integral of the
real part over $\omega$ from 0 to $+\infty$.

\subsection{Integration over frequency }

To carry out the integration over $\omega$, we need to make analytic
continuation for the frequency to complex plane and calculate the
residues of singularities. The denominator $D(\omega,k_{T})$ in Eq.
(\ref{eq:int-omega-k}) is a quartic polynomial of $\omega$. For
a fixed $k_{T}$, $D(\omega,k_{T})$ has four roots in the complex
plane, each of which gives a pole of the integrand. In high energy
heavy ion collisions, the Ohm conductivity $\sigma$ is much larger
than the chiral magnetic conductivity $\sigma_{\chi}$, so it is reasonable
to treat $\sigma_{\chi}$ as a perturbation. 

Now we deal with the poles of the integrand in Eq. (\ref{eq:int-omega-k}).
To this end, we need to find the roots of the equation $D(\omega,k_{T})=0$.
For a fixed value of $k_{T}$, we assume the solutions take the following
form 
\begin{equation}
\omega=\omega_{0}+\sigma_{\chi}c_{1}+\sigma_{\chi}^{2}c_{2}+\cdots,\label{eq:omega-expand}
\end{equation}
where the zero-th order value $\omega_{0}$ denote the roots of the
equation $L(\omega,k_{T})=0$ and are given by 
\begin{eqnarray}
\omega_{\pm} & \equiv & iv\gamma\frac{1}{2}\left[v\gamma\sigma\pm\sqrt{(v\gamma\sigma)^{2}+4k_{T}^{2}}\right].\label{eq:omega-pm}
\end{eqnarray}
We see that the zero-th order solutions are doublet. We insert Eq.
(\ref{eq:omega-expand}) into $D(\omega,k_{T})$ and expand in powers
of $\sigma_{\chi}$. When implementing $\omega_{0}=\omega_{\pm}$
in $D(\omega,k_{T})$, the zero-th and first order terms in $\sigma_{\chi}$
are vanishing. The coefficient $c_{1}$ appears in the $\sigma_{\chi}^{2}$
term and can be determined by the condition that it vanishes. Implementing
the values of $c_{1}$, we can determine $c_{2}$ from the vanishing
of $\sigma_{\chi}^{3}$ term. Putting them together, we obtain the
roots in the following form 
\begin{eqnarray}
\omega_{s_{1}s_{2}} & \equiv & \omega_{s_{1}}+s_{2}\sigma_{\chi}c_{s_{1}}^{(1)}+\sigma_{\chi}^{2}c_{s_{1}}^{(2)},\qquad(s_{1},s_{2}=\pm1),\label{eq:sulutions}
\end{eqnarray}
where $c_{s_{1}}^{(1)}$ and $c_{s_{1}}^{(2)}$ are given by 
\begin{eqnarray}
c_{s}^{(1)} & = & \frac{v\gamma^{2}\sqrt{(v\gamma\sigma)^{2}+2v^{2}k_{T}^{2}+s(v\gamma\sigma)\sqrt{(v\gamma\sigma)^{2}+4k_{T}^{2}}}}{\sqrt{2}\sqrt{(v\gamma\sigma)^{2}+4k_{T}^{2}}},\nonumber \\
c_{s}^{(2)} & = & -s\frac{iv\gamma^{3}k_{T}^{2}(2-v^{2})}{\left[(v\gamma\sigma)^{2}+4k_{T}^{2}\right]^{3/2}}.
\end{eqnarray}
The polynomial $D(\omega,k_{T})$ can thus be expressed in terms of
these four roots in (\ref{eq:sulutions}), 
\begin{eqnarray}
D(\omega,k_{T}) & = & \frac{1}{(v\gamma)^{4}}\prod_{s_{1},s_{2}=\pm}(\omega-\omega_{s_{1}s_{2}}).
\end{eqnarray}
It is easy to verify that $\omega_{++}$ and $\omega_{+-}$ are located
in the upper half complex plane and $\omega_{--}$ are located in
the lower half one, while $\omega_{-+}$ is located in the lower half
plane when $\sigma_{\chi}<k_{T}$. For a relativistic particle in
heavy ion collisions, the condition $\sigma_{\chi}<k_{T}$ is satisfied
in most cases \cite{Tuchin:2014iua}. So we treat $\omega_{-+}$ as
a pole located in the lower half plane. 

To the linear order in $\sigma_{\chi}$, the differences between two
poles in the upper and lower half plane are 
\begin{eqnarray}
\Delta\omega_{+} & = & \omega_{++}-\omega_{+-}\approx2\sigma_{\chi}c_{+}^{(1)}\rightarrow2\sigma_{\chi}v\gamma^{2},\;\;\mathrm{for}\: k_{T}=0\nonumber \\
\Delta\omega_{-} & = & \omega_{-+}-\omega_{--}\approx2\sigma_{\chi}c_{-}^{(1)}\rightarrow0,\;\;\mathrm{for}\: k_{T}=0
\end{eqnarray}
We see that $\Delta\omega_{+}$ ($\Delta\omega_{-}$) is non-vanishing
(vanishing) at $k_{T}=0$. For the imaginary part in Eq. (\ref{eq:omega-pm}),
we see $|\omega_{+}|\geq(v\gamma)^{2}\sigma$, i.e. it has non-zero
lower bound but $|\omega_{-}|\geq0$ has zero bound, where the equality
holds at $k_{T}=0$ for both cases. The poles in upper (lower) half
plane whose imaginary part is $\omega_{+}(\omega_{-})$ give the advanced
(retarded) solution with $vt-z<0$ ($vt-z>0$). Such a difference
in the imaginary part makes the advanced solution more suppressed
than the retarded one at relativitic limit with $\gamma\gg1$. 

Then we can carry out the integration over $\omega$ by contour integration.
For the advanced (retarted) region $vt<z$ ($vt>z$) , we need to
close the contour in the upper (lower) half plane and pick up two
poles $\omega_{+,\pm}$ ($\omega_{-,\pm}$). The residues of $D^{-1}(\omega,k_{T})$
at poles are given by 
\begin{eqnarray}
R_{s_{1}s_{2}}(k_{T}) & \equiv & \lim_{\omega\rightarrow\omega_{s_{1}s_{2}}}\frac{\omega-\omega_{s_{1}s_{2}}}{D(\omega,k_{T})}\approx\frac{(v\gamma)^{4}}{2\sigma_{\chi}c_{s_{1}}^{(1)}(\delta\omega)^{2}}\left(s_{2}-s_{1}\frac{2\sigma_{\chi}c_{s_{1}}^{(1)}}{\delta\omega}\right),\label{eq:residues-1}
\end{eqnarray}
where $\delta\omega\equiv\omega_{+}-\omega_{-}=iv\gamma\sqrt{(v\gamma\sigma)^{2}+4k_{T}^{2}}$
is the difference between two roots in Eq. (\ref{eq:omega-pm}). Thus
the integration over $\omega$ can be done by applying the residue
theorem, 
\begin{eqnarray}
\mathbf{B}\left(t,\mathbf{x}\right) & = & -i\theta(\frac{z}{v}-t)Q\int\frac{dk_{T}}{2\pi}\sum_{s=\pm}\mathbf{B}^{\prime}(\omega_{+,s},k_{T})R_{+,s}(k_{T})\nonumber \\
 &  & +i\theta(t-\frac{z}{v})Q\int\frac{dk_{T}}{2\pi}\sum_{s=\pm}\mathbf{B}^{\prime}(\omega_{-,s},k_{T})R_{-,s}(k_{T}),\label{eq:mag-residue}
\end{eqnarray}
where the terms of $\theta(\frac{z}{v}-t)$ and $\theta(t-\frac{z}{v})$
correspond to advanced and retarded contributions respectively.

\subsection{Algebraic expressions for magnetic fields}

After carrying out the $k_{T}$ integral, we obtain an algebraic expression
for the tangential component $B_{\phi}$ in the leading order in $\sigma_{\chi}$, 

\begin{eqnarray}
B_{\phi}\left(t,\mathbf{x}\right) & = & \frac{Q}{4\pi}\cdot\frac{v\gamma x_{T}}{\Delta^{3/2}}\left(1+\frac{\sigma v\gamma}{2}\sqrt{\Delta}\right)e^{A}.\label{eq:mag-phi}
\end{eqnarray}
Here we have defined symbols $\Delta\equiv\gamma^{2}(vt-z)^{2}+x_{T}^{2}$
and $A\equiv(\sigma v\gamma/2)[\gamma(vt-z)-\sqrt{\Delta}]$ (note
that $A<0$). It is easy to verify that $B_{\phi}$ in Eq. (\ref{eq:mag-phi})
recovers the formula from Lienard-Wiechert potentials when $\sigma=0$.
We note that such a form of $B_{\phi}$ was first given in Ref. \cite{Gursoy:2014aka}.
The linear order contribution in $\sigma_{\chi}$ is absent in $B_{\phi}$.
This means that the chiral magnetic effect characterized by $\sigma_{\chi}$
does not play a major role in the tangential component. However the
major correction is from electric conductivity. 

For the radial and longitudinal components, we now give following
simple algebraic expressions in the leading order in $\sigma_{\chi}$,
\begin{eqnarray}
B_{r}\left(t,\mathbf{x}\right) & = & -\sigma_{\chi}\frac{Q}{8\pi}\cdot\frac{v\gamma^{2}x_{T}}{\Delta^{3/2}}\left[\gamma(vt-z)+A\sqrt{\Delta}\right]e^{A},\nonumber \\
B_{z}\left(t,\mathbf{x}\right) & = & \sigma_{\chi}\frac{Q}{8\pi}\cdot\frac{v\gamma}{\Delta^{3/2}}\left[\gamma^{2}(vt-z)^{2}\left(1+\frac{\sigma v\gamma}{2}\sqrt{\Delta}\right)+\Delta\left(1-\frac{\sigma v\gamma}{2}\sqrt{\Delta}\right)\right]e^{A}.\label{eq:mag-rz}
\end{eqnarray}
We see that they are proportional to $\sigma_{\chi}$. Previous studies
have shown that the electric conducting effect will never generate
$B_{r}$ and $B_{z}$, so these non-vanishing components are the result
of the chiral magnetic effect. This can be easily understood: a moving
charge produces magnetic fields in the tangential direction, which
then turns into a tangential current due to the chiral magnetic effect
and finally generates $B_{r}$ and $B_{z}$. 

We now make a few comments about advanced and retarted contributions
in Eqs. (\ref{eq:mag-phi},\ref{eq:mag-rz}). We see that $\theta(\frac{z}{v}-t)$
and $\theta(t-\frac{z}{v})$ which characterize the advanced and retarded
contributions disappear, the reason is that the rest expressions apart
from the theta-functions are identical in both contributions, therefore
we can combine them as $\theta(\frac{z}{v}-t)+\theta(t-\frac{z}{v})=1$.
The presence of the factor $e^{A}$ shows that the advanced contribution
is suppressed exponentially relative to the retarded one, since the
$\sigma\gamma^{2}(vt-z)$ part in $A$ is negative (positive) for
the advanced (retarded) parts. 

One can verify that the higher order corrections to $B_{\phi}$, $B_{r}$
and $B_{z}$ are all of $O(\sigma_{\chi}^{2})$. At very late time,
one can see from Eqs. (\ref{eq:mag-phi},\ref{eq:mag-rz}) that the
fields decay in time as $B_{\phi,r}\sim1/t^{2}$ and $B_{z}\sim1/t$.

\subsection{Relativistic limit}

In this subsection, we consider the relativistic limit with $v\sim1$
and $\gamma\gg1$. Eq. (\ref{eq:denominator}) becomes, 
\begin{eqnarray}
L(\omega,k_{T}) & \approx & i\sigma\omega-k_{T}^{2},\nonumber \\
D(\omega,k_{T}) & \approx & \left(i\sigma\omega-k_{T}^{2}\right)^{2}-\sigma_{\chi}^{2}\omega^{2}-\sigma_{\chi}^{2}k_{T}^{2}.\label{eq:denominator-1}
\end{eqnarray}
The $D=0$ has two roots for a given $k_{T}$, 
\begin{equation}
\omega_{\pm}=-\frac{i\sigma}{\sigma^{2}+\sigma_{\chi}^{2}}k_{T}^{2}\pm\frac{\sigma_{\chi}}{\sigma^{2}+\sigma_{\chi}^{2}}k_{T}\sqrt{k_{T}^{2}-(\sigma^{2}+\sigma_{\chi}^{2})}.
\end{equation}
If we focus on the region $k_{T}\gg\sigma,\sigma_{\chi}$, these can
be simplified to 
\begin{equation}
\omega_{\pm}=\frac{k_{T}^{2}}{i\sigma\pm\sigma_{\chi}},
\end{equation}
These two roots are all under the real axis, which means that the
advanced solution is vanishing. So the contour integration over $\omega$
in the lower half plane pickes up these two poles at $\omega_{\pm}$.
We can carry out the integration of Bessel functions 
\begin{eqnarray}
\int_{0}^{\infty}dkk^{2}\exp[-iak^{2}]J_{1}(kb) & = & -\frac{b}{4a^{2}}\exp\left[i\frac{b^{2}}{4a}\right],\nonumber \\
\int_{0}^{\infty}dkk\exp[-iak^{2}]J_{0}(kb) & = & -\frac{i}{2a}\exp\left[i\frac{b^{2}}{4a}\right].
\end{eqnarray}
Finally we obtain the analytical expressions for the magnetic fields
\begin{eqnarray}
B_{r}(t,\mathbf{x}) & = & \theta(t-z)Q\frac{x_{T}}{8\pi(t-z)^{2}}\exp\left[-\frac{\sigma x_{T}^{2}}{4(t-z)}\right]\nonumber \\
 &  & \times\left\{ \sigma\sin\left[\frac{\sigma_{\chi}x_{T}^{2}}{4(t-z)}\right]-\sigma_{\chi}\cos\left[\frac{\sigma_{\chi}x_{T}^{2}}{4(t-z)}\right]\right\} ,\nonumber \\
B_{\phi}(t,\mathbf{x}) & = & \theta(t-z)Q\frac{x_{T}}{8\pi(t-z)^{2}}\exp\left[-\frac{\sigma x_{T}^{2}}{4(t-z)}\right]\nonumber \\
 &  & \times\left\{ \sigma\cos\left[\frac{\sigma_{\chi}x_{T}^{2}}{4(t-z)}\right]+\sigma_{\chi}\sin\left[\frac{\sigma_{\chi}x_{T}^{2}}{4(t-z)}\right]\right\} ,\nonumber \\
B_{z}(t,\mathbf{x}) & = & \theta(t-z)Q\frac{1}{4\pi(t-z)}\exp\left[-\frac{\sigma x_{T}^{2}}{4(t-z)}\right]\nonumber \\
 &  & \times\left\{ -\sigma\sin\left[\frac{\sigma_{\chi}x_{T}^{2}}{4(t-z)}\right]+\sigma_{\chi}\cos\left[\frac{\sigma_{\chi}x_{T}^{2}}{4(t-z)}\right]\right\} .\label{eq:b-rel-lim}
\end{eqnarray}
We see that only $B_{\phi}$ is non-vanishing at $\sigma_{\chi}=0$.
For a point charge moving in the opposite direction, $v\sim-1$, the
magnetic fields {[}up to $\theta(t+z)${]} can be obtained from Eq.
(\ref{eq:b-rel-lim}) by a rotation along any radial axis on the transverse
plane at $z=0$ . In this case, $B_{\phi}$ and $B_{z}$ change their
signs but $B_{r}$ does not. 

One can verify that these fields satisfy the Maxwell equations (\ref{eq:field-eq-sigma}).
In the same way, we can also derive analytic formula for electric
fields in the relativistic limit but the expressions are much more
complicated than magnetic fields.

\section{Electric Fields of a moving charge}

\textbf{\Large \label{sec:elect}}In this section, we will derive
the analytical expression for electric fields in a medium with both
Ohm conductivity and chiral magnetic conductivity. Same as in Section
\ref{2.-Magnetic-Fields}, we consider that a charged particle moves
in the third direction. Following the procedure similar to Section
\ref{2.-Magnetic-Fields}, we obtain 
\begin{eqnarray}
\left(\begin{array}{c}
E_{x}\\
E_{y}\\
E_{z}
\end{array}\right)(\omega,\mathbf{k}) & = & 2\pi iQ\omega\frac{\delta(\omega-k_{z}v)}{L^{2}-\sigma_{\chi}^{2}k^{2}}\nonumber \\
 &  & \times\left[\frac{L+\sigma_{\chi}^{2}}{\omega+i\sigma}\left(\begin{array}{c}
k_{x}\\
k_{y}\\
k_{z}
\end{array}\right)+i\sigma_{\chi}v\left(\begin{array}{c}
k_{y}\\
-k_{x}\\
0
\end{array}\right)-vL\left(\begin{array}{c}
0\\
0\\
1
\end{array}\right)\right].\label{eq:e-mom}
\end{eqnarray}
With Eq. (\ref{eq:b-mom}) for $\mathbf{B}$ and Eq. (\ref{eq:e-mom})
for $\mathbf{E}$ we can verify that the Maxwell equtions are really
satisfied. 

When transforming back to coordinate space, we follow the same procedure
as in Section \ref{2.-Magnetic-Fields} and get a form for electric
fields similar to Eq. (\ref{eq:int-omega-k}) for magnetic fields,
\begin{eqnarray}
\mathbf{E}\left(t,\mathbf{x}\right) & = & -\frac{Q}{v}\int\frac{d\omega dk_{T}}{(2\pi)^{2}}\:\frac{\mathbf{E}^{\prime}(\omega,k_{T})}{D(\omega,k_{T})}.\label{eq:electric}
\end{eqnarray}
In the cylindrical coordinate, $\mathbf{E}^{\prime}(\omega,k_{T})$
is given by 
\begin{eqnarray}
\left(\begin{array}{c}
E_{r}^{\prime}\\
E_{\phi}^{\prime}\\
E_{z}^{\prime}
\end{array}\right)(\omega,k_{T}) & = & ik_{T}\omega\mathrm{e}^{-i\omega(t-z/v)}\left[\frac{L(\omega,k_{T})+\sigma_{\chi}^{2}}{i\omega-\sigma}\left(\begin{array}{c}
k_{T}J_{1}(k_{T}x_{T})\\
0\\
-i(\omega/v)J_{0}(k_{T}x_{T})
\end{array}\right)\right.\nonumber \\
 &  & \left.+\left(\begin{array}{c}
0\\
-\sigma_{\chi}vk_{T}J_{1}(k_{T}x_{T})\\
vL(\omega,k_{T})J_{0}(k_{T}x_{T})
\end{array}\right)\right].\label{eq:e-cyl}
\end{eqnarray}
But the difference from the case of magnetic fields is: besides the
four poles in $1/D(\omega,k_{T})$, there is an additional pole in
the lower half plane from the first term $\sim1/(\omega+i\sigma)$
as shown in Eq. (\ref{eq:e-mom}). 

From Maxwell equations, we can obtain $E_{\phi}$ from $B_{r}$ instantly
($E_{\phi}=-vB_{r}$), 
\begin{eqnarray}
E_{\phi} & = & \sigma_{\chi}\frac{Q}{8\pi}\frac{v^{2}\gamma^{2}x_{T}}{\Delta^{3/2}}\left[\gamma(vt-z)+A\sqrt{\Delta}\right]e^{A}.\label{eq:e-phi}
\end{eqnarray}
Generally the integration over $k_{T}$ in $E_{r,z}$ cannot be worked
out analytically due to the term $1/(i\omega-\sigma)$ in Eq. (\ref{eq:e-cyl}).
However, at relativitic limit $\gamma\gg1$, this can be done and
we can obtain algebraic expression for $E_{r,z}$, 
\begin{eqnarray}
E_{r} & = & \frac{Q}{4\pi}\left\{ \frac{\gamma x_{T}}{\Delta^{3/2}}\left(1+\frac{\sigma v\gamma}{2}\sqrt{\Delta}\right)-\frac{\sigma}{vx_{T}}e^{-\sigma(t-z/v)}\left[1+\frac{\gamma(vt-z)}{\sqrt{\Delta}}\right]\right\} e^{A},\nonumber \\
E_{z} & = & \frac{Q}{4\pi}\left\{ -e^{A}\frac{1}{\Delta^{3/2}}\left[\gamma(vt-z)+A\sqrt{\Delta}+\frac{\sigma\gamma}{v}\Delta\right]+\frac{\sigma^{2}}{v^{2}}e^{-\sigma(t-z/v)}\Gamma(0,-A)\right\} ,\label{eq:erz}
\end{eqnarray}
where $\Gamma(0,-A)$ is the incomplete gamma function defined as
$\Gamma(a,z)=\int_{z}^{\infty}dt\, t^{a-1}\exp(-t)$. We have checked
in numerical calculations that the result of Eqs. (\ref{eq:e-phi},\ref{eq:erz})
is a good approximation to the exact result for the scale of heavy
ion collisions. We have also checked that electric and magnetic fields
in Eqs. (\ref{eq:e-phi},\ref{eq:erz}) and Eqs. (\ref{eq:mag-phi},\ref{eq:mag-rz})
satisfy the Maxwell equations (\ref{eq:field-eq-sigma}) in a good
accuracy for the scale of heavy ion collisions. 

In the leading order in $\sigma_{\chi}$, we see in Eqs. (\ref{eq:e-phi},\ref{eq:erz})
that $E_{\phi}$ is proportional to $\sigma_{\chi}$ while $E_{r,z}$
are independent of $\sigma_{\chi}$. The higher order contributions
to $E_{\phi}$, $E_{r}$ and $E_{z}$ are all of $O(\sigma_{\chi}^{2})$. 

\begin{figure}
\caption{\label{fig:point-eb}The electromagnetic fields at $\mathbf{x}=(0,0,0)$
fm produced by a point charge (proton) of 100 GeV which are located
at $\mathbf{x}=(6,0,0)$ fm and moving along $\mathbf{e}_{z}$. We
choose following values for conductivities $\sigma=5.8$ MeV and $\sigma_{\chi}=1.5$
MeV. }

\includegraphics[scale=0.8]{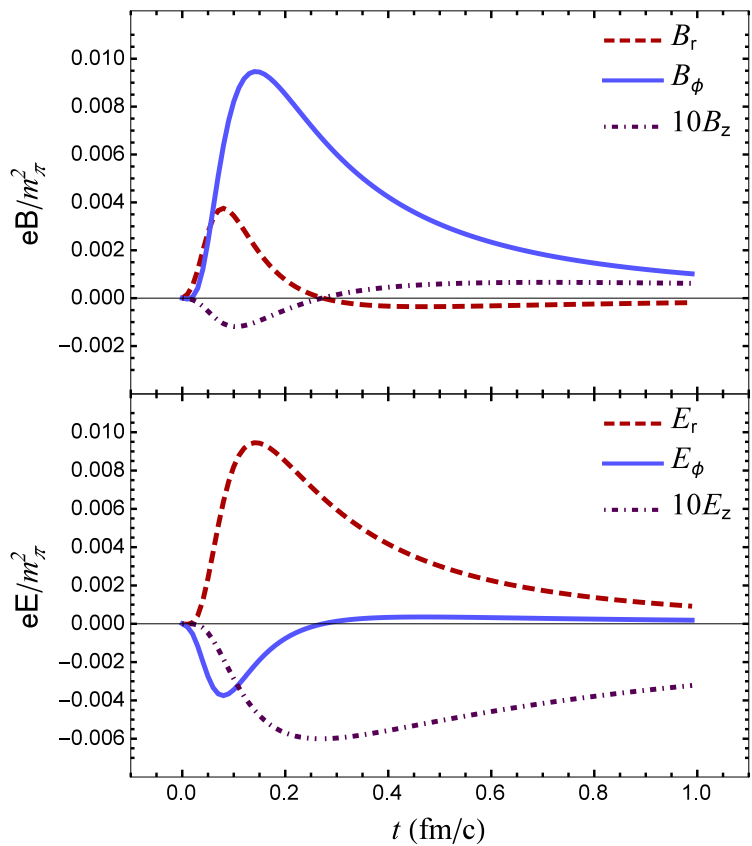}
\end{figure}

\begin{figure}
\caption{\label{fig:collision}The geometry of two colliding nuclei in the
transverse plane at $z=0$. One nucleus at $(b/2,0)$ in the transverse
plane is moving along $\mathbf{e}_{z}$, while another nucleus at
$(-b/2,0)$ is moving along $-\mathbf{e}_{z}$. The points $P$ and
$Q$ in the transverse plane are two typical points at which $\mathbf{B}$
and $\mathbf{E}$ will be calculated. }

\includegraphics[scale=0.4]{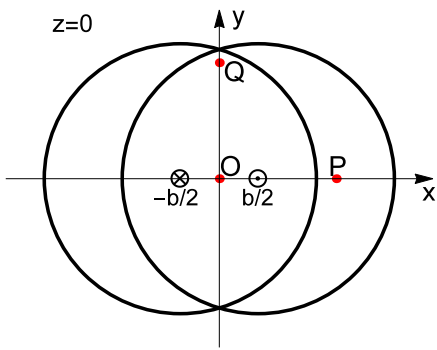}
\end{figure}

\begin{figure}
\caption{\label{fig:time-evo2}Time evolution of $B_{y}$ and $E_{y}$ in Au+Au
collisions at $\sqrt{s}=$200 GeV and $\mathbf{x}=(0,6,0)$ fm for
three cases: (a) Lienard-Wiechert potential ($\sigma=\sigma_{\chi}=0$);
(b) with $\sigma$ ($\sigma\neq0$ and $\sigma_{\chi}=0$); (c) with
$\sigma$ and $\sigma_{\chi}$ ($\sigma\neq0$ and $\sigma_{\chi}\neq0$).
The $x$ and $z$ components are vanishing, $B_{x,z}\approx0$ and
$E_{x,z}\approx0$. }

\includegraphics[scale=0.5]{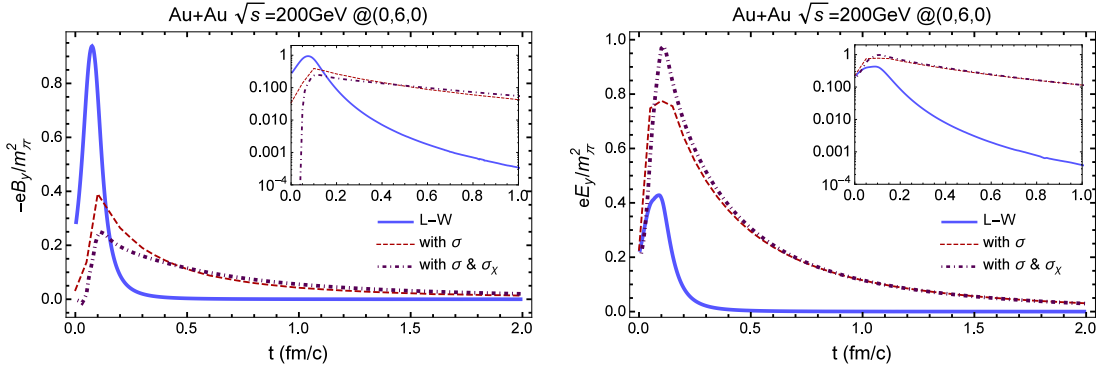}
\end{figure}

\begin{figure}
\caption{\label{fig:time-evo1}The time evolution of $\mathbf{B}$ and $\mathbf{E}$
in Au+Au collisions at $\sqrt{s}=$200 GeV and $\mathbf{x}=(6,0,0)$
fm for three cases: (a) Lienard-Wiechert potential (L-W, $\sigma=\sigma_{\chi}=0$);
(b) with $\sigma$ ($\sigma\neq0$ and $\sigma_{\chi}=0$); (c) with
$\sigma$ and $\sigma_{\chi}$ ($\sigma\neq0$ and $\sigma_{\chi}\neq0$). }

\includegraphics[scale=0.5]{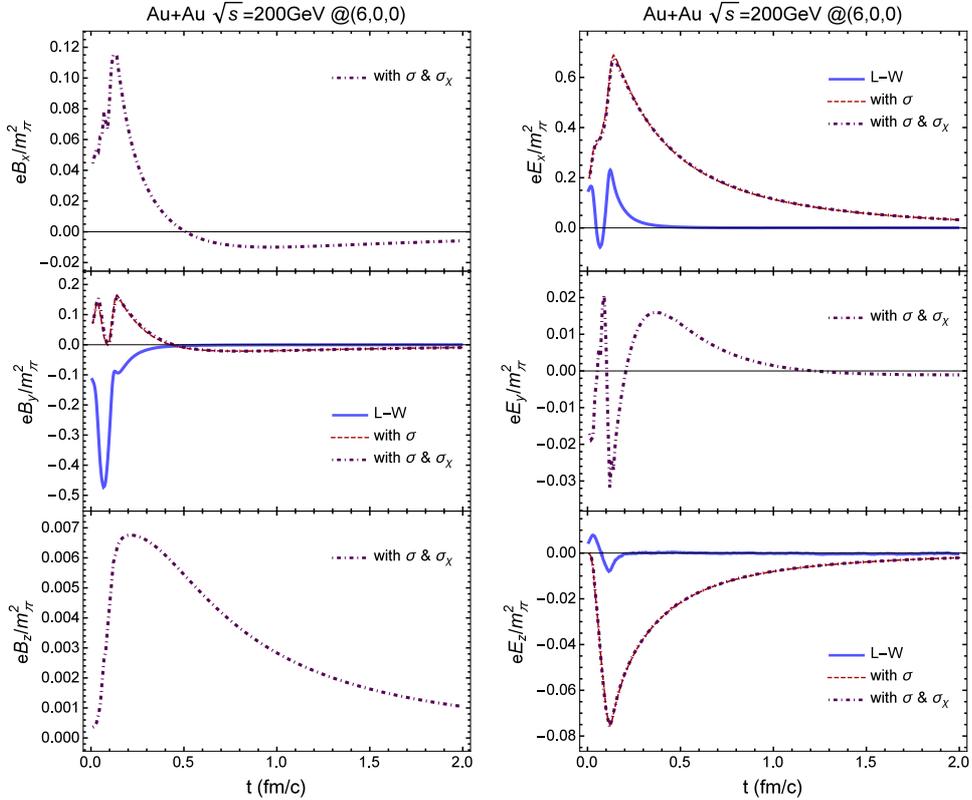}
\end{figure}

\begin{figure}
\caption{\label{fig:contour}The contour plots for electric (upper panel) and
magnetic (lower panel) fields in the transverse plane of $z=0$ at
$t=2$ fm/c and $\sqrt{s}=200$ GeV in Au+Au collisions. The two colliding
nuclei are shown in two red dashed circles. }
\includegraphics[scale=0.6]{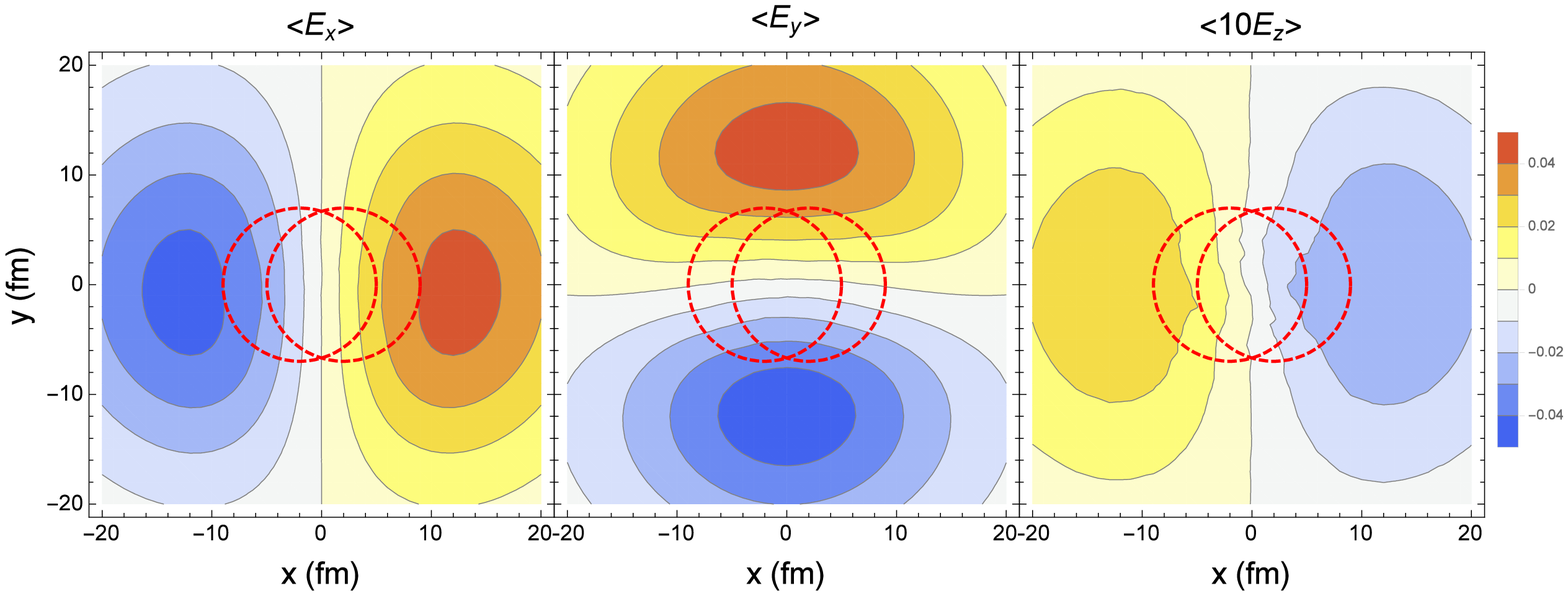}

\includegraphics[scale=0.6]{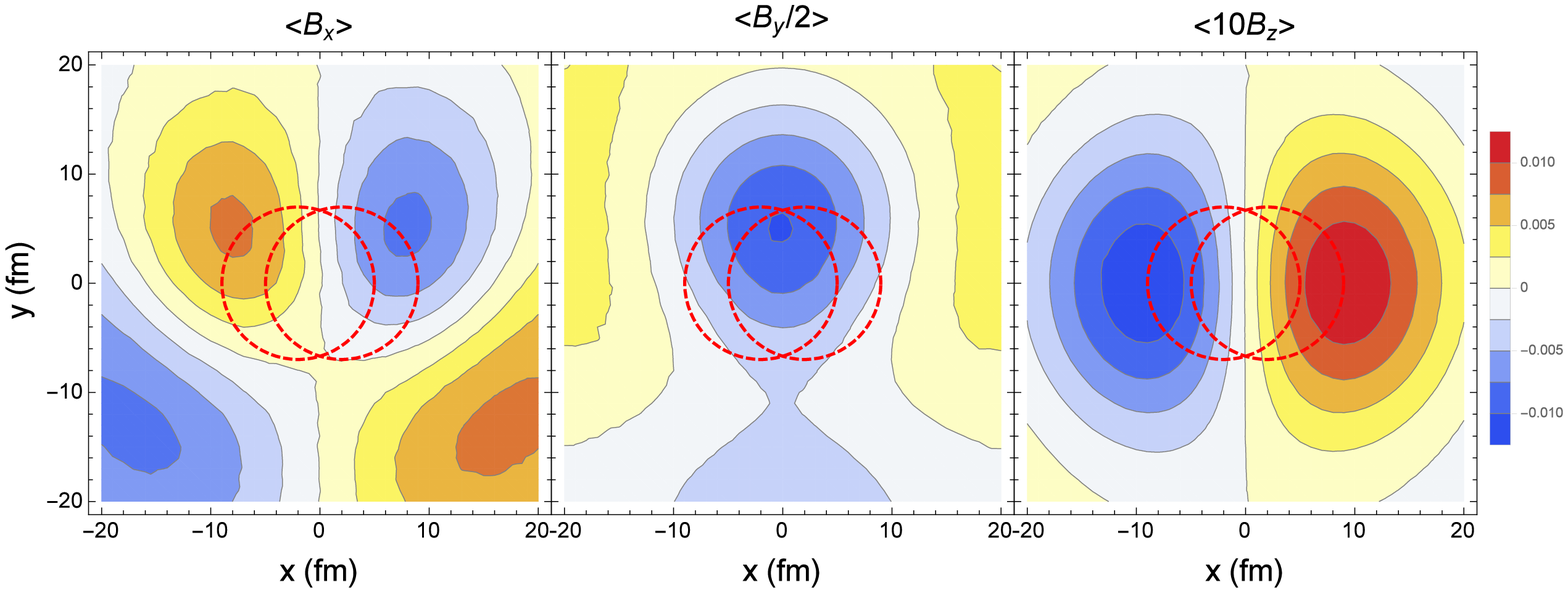}
\end{figure}

\begin{figure}
\caption{\label{fig:vector}The 2-dimensional vector fields for transverse
components $\mathbf{B}_{T}$ and $\mathbf{E}_{T}$ in the transverse
plane of $z=0$ at $t=2$ fm/c and $\sqrt{s}=200$ GeV in Au+Au collisions. }

\includegraphics[scale=0.6]{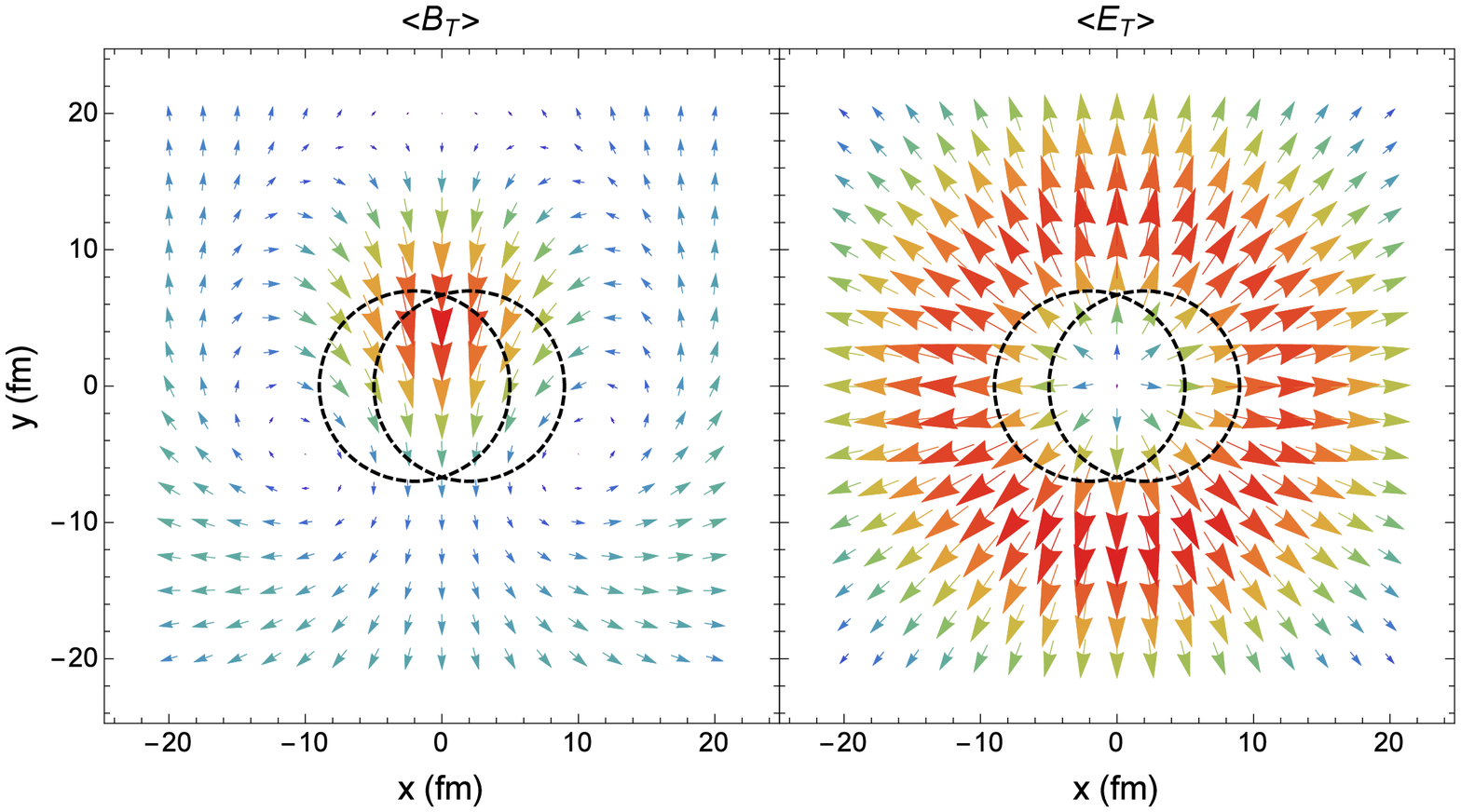}
\end{figure}

\section{Numerical results for electromagnetic fields in heavy-ion collisons}

\label{sec:numerical}In this section we will give numerical results
for $\mathbf{B}$ and $\mathbf{E}$ from Eqs. (\ref{eq:int-omega-k},\ref{eq:electric}).
The source terms are given by the configuration that two nuclei collide
with an impact parameter, which is a convolution of the point charge/current
density in the form of Eq. (\ref{eq:source-x-space}) with the charge
distribution of nuclei. 

Fig. (\ref{fig:point-eb}) shows $\mathbf{B}$ and $\mathbf{E}$ as
functions of time at $\mathbf{x}=(0,0,0)$ fm produced by a point
charge (proton) of 100 GeV located at $(6,0,0)$ fm and moving along
$\mathbf{e}_{z}$. We see that the magnitude of $B_{\phi}$ is larger
than $B_{r}$ almost all the time and $B_{z}$ is much smaller than
$B_{r}$ and $B_{\phi}$. The non-vanishing $B_{r}$ and $B_{z}$
is due to the chiral magnetic effect or $\sigma_{\chi}\neq0$. We
also see that the magnitude of $E_{r}$ is larger than $E_{\phi}$
(just opposite to the magnetic field) and $E_{z}$. All field components
of $\mathbf{B}$ and $\mathbf{E}$ are damped as the time goes on. 

We show in Fig. (\ref{fig:collision}) the geometry of two colliding
nuclei in peripheral collisions with the impact parameter $b$. The
global magnetic field of this configuration is along $-\mathbf{e}_{y}$.
In the numerical calculation of $\mathbf{B}$ and $\mathbf{E}$, we
choose $b=4$ fm for Au+Au collisions at $\sqrt{s}=200$ GeV. We use
UrQMD to simulate the space-time and momentum configurations of charged
particles in Au+Au collisions in the case of Lienard-Wiechert potential.
After the collisions, the spectator nucleons which do not collide
fly by freely while participant nucleons are stopped to produce new
particles. Participant nucleons will treated differently in the cases
of non-vanshing medium effects with $\sigma$ and $\sigma_{\chi}$:
the rapidity distribution of charged particles produced by participant
nucleons has to be modified. In our calculations, we adopt the rapidity
distribution in Ref. \cite{Gursoy:2014aka}. 

We show in Figs. (\ref{fig:time-evo2},\ref{fig:time-evo1}) the time
evolution of $\mathbf{B}$ and $\mathbf{E}$ in Au+Au collisions at
$\sqrt{s}=$200 GeV and at two points $(6,0,0)$ fm (the point $P$
in Fig. (\ref{fig:collision})) and $(0,6,0)$ fm (the point $Q$
in Fig. (\ref{fig:collision})). We consider three cases: (a) Lienard-Wiechert
potential ($\sigma=\sigma_{\chi}=0$, blue solid lines); (b) with
only $\sigma$ ($\sigma\neq0$ and $\sigma_{\chi}=0$, red dashed
lines); (c) with both $\sigma$ and $\sigma_{\chi}$ ($\sigma\neq0$
and $\sigma_{\chi}\neq0$, magenta dash-dotted lines). 

In Fig. (\ref{fig:time-evo2}) we give the time evolution of $B_{y}$
and $E_{y}$ at the point $\mathbf{x}=(0,6,0)$ fm or the point $Q$.
The $x$ and $z$ components are vanishing, $B_{x,z}\approx0$ and
$E_{x,z}\approx0$ due to that $OQ$ is along the direction of global
orbital angular momentum or global magnetic field. The effect of $\sigma_{\chi}$
on $B_{y}$ and $E_{y}$ is small at late time. 

In Fig. (\ref{fig:time-evo1}), we see that $B_{x}$, $B_{z}$ and
$E_{y}$ are mainly controlled by $\sigma_{\chi}$, i.e. they are
vanishing at $\sigma_{\chi}=0$. It is interesting to see that $B_{y}$
has different signs from L-W and from $\sigma\neq0$ in very short
time from the collision moment. The reason is that $B_{y}$ with L-W
is from spectators moving apart rapidly so it is along $-\mathbf{e}_{y}$
and decays quickly in time, but $B_{y}$ with non-vanishing $\sigma$
is dominated by the conducting current and lasts longer than the L-W
contribution. 

The contour plots for electric and magnetic fields in the transverse
plane of $z=0$ are shown in Figs. (\ref{fig:contour}). The time
is set to at $t=2$ fm/c. We see that the magnitudes of $x,z$ components
of electric fields $|E_{x,z}(t,x,y,z)|$ are symmetric for flipping
the signs of their arguments $x$ and $y$. The symmetry is partially
broken for $|E_{y}(t,x,y,z)|$ and $|B_{x,y}(t,x,y,z)|$ due to $\sigma_{\chi}$:
they are symmetric for flipping the sign of $x$ but not for $y$,
while $\left|B_{z}(t,x,y,z)\right|$ preserves the symmetry for flipping
the signs of $x$ and $y$. The field configuration can be more clearly
seen in Fig. (\ref{fig:vector}), where the transverse components
are shown in two-dimension vectors. We see that $\mathbf{E}_{T}$
is more symmetric than $\mathbf{B}_{T}$ in the transverse plane.
A magnetic field along $-\mathbf{e}_{y}$ can also be clearly seen
near the origin $(0,0,0)$. It is obvious that $|B_{y}(t,x,y,z)|\neq|B_{y}(t,x,-y,z)|$. 

\begin{figure}
\caption{\label{fig:asymmetry-b}Illustration of the asymmetry of the magnetic
fields at non-vanishing $\sigma_{\chi}$. Two positive point charges
at $(\pm a,0,0)$ move in $\pm\mathbf{e}_{z}$ direction. (a) Azimuthal
components; (b) Radial components. The azimuthal components are symmetric
at symmetric points $(0,\pm b,0)$, while the radial components have
opposite signs. }

\includegraphics[scale=0.4]{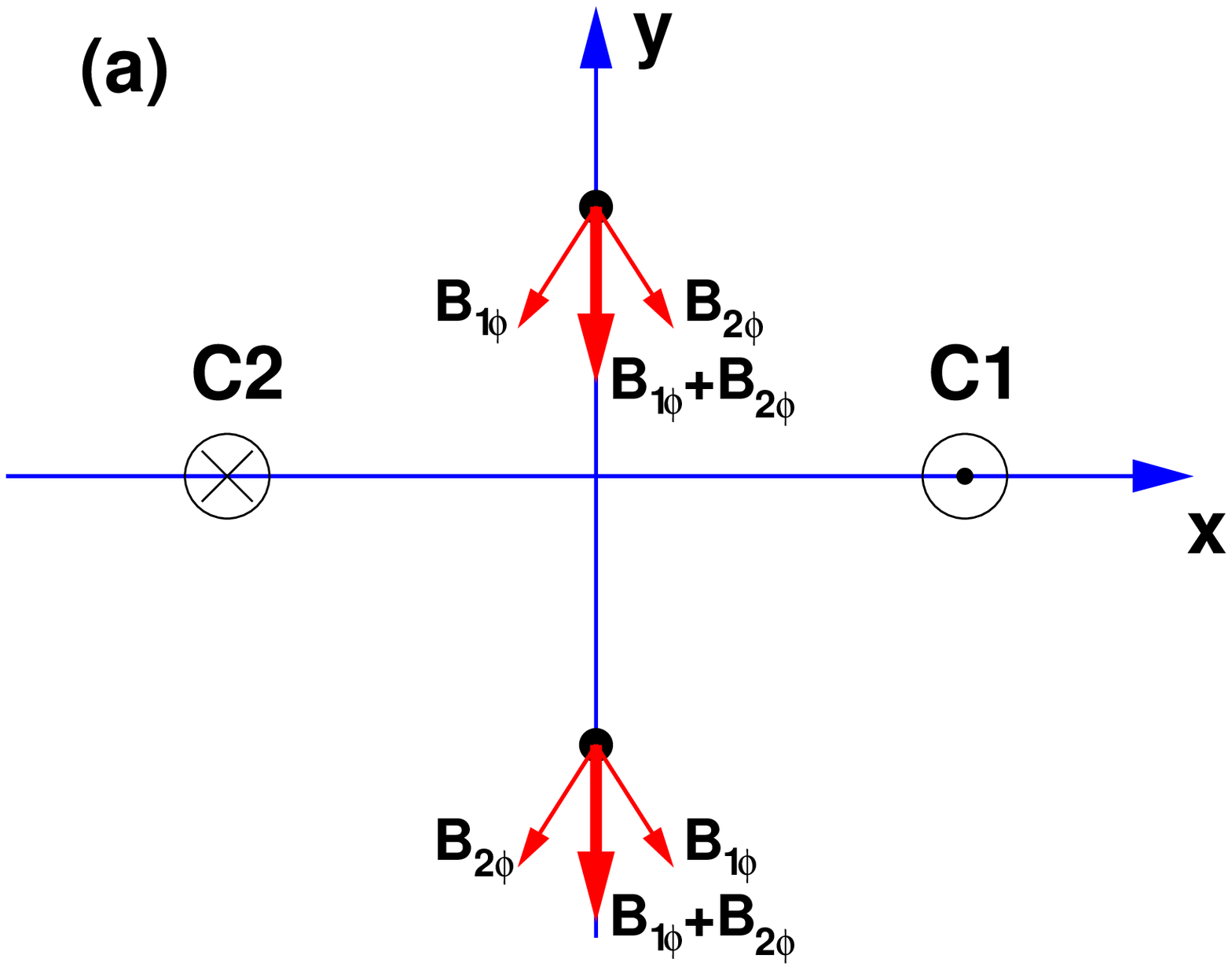}\includegraphics[scale=0.4]{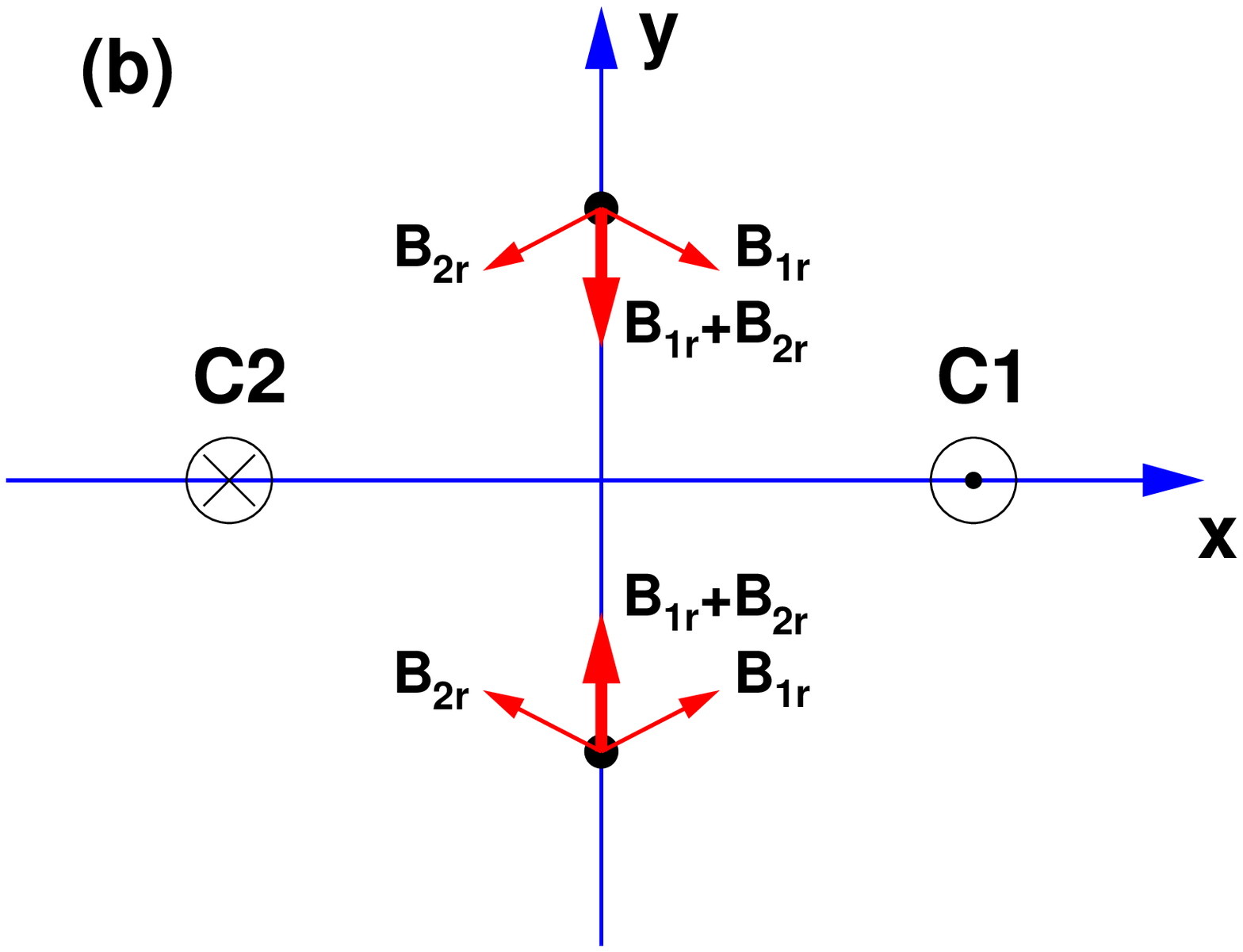}
\end{figure}

The asymmetry in Figs. (\ref{fig:contour},\ref{fig:vector}) can
be easily understood from non-vanishing $B_{r}$ resulting from $\sigma_{\chi}$.
Suppose one positive charge is located at $(a,0,0)$ fm and moving
along $-\mathbf{e}_{z}$, while the other one is located at $(-a,0,0)$
fm and moving along $\mathbf{e}_{z}$, see Fig. \ref{fig:asymmetry-b}.
We can compare the magnetic fields at two points, $(0,b,0)$ fm and
$(0,-b,0)$ fm. For simplicity we assume the relativistic limit and
use Eq. (\ref{eq:b-rel-lim}), where we observe that $B_{r}$ does
not change the sign when flipping the velocity direction. Therefore
the direction of radial components of the magnetic fields from two
oppositely moving charges at upper point $(0,b,0)$ fm is opposite
to that at lower point $(0,-b,0)$. But azimuthal components have
the same directions and magnitudes at upper and lower point. Thus
the total magnetic fields, or the vector sums of radial and azimuthal
components, have different magnitude at two symmetric points with
respect to the $x$-axis. 

\section{Summary}

\label{sec:summary}We have derived analytic expressions for electric
and magnetic fields produced by a point charge in a conducting medium
with the electric conductivity $\sigma$ and the chiral magnetic conductivity
$\sigma_{\chi}$. We used the method of the Green function under the
condition $\sigma\gg\sigma_{\chi}$. We have given in Eqs. (\ref{eq:mag-rz},\ref{eq:e-phi},\ref{eq:erz})
for the first time the algebraic expressions for electric and magnetic
fields as functions of space-time without any integrals. Numerical
results show that these algebraic results work very well for values
of $\sigma_{\chi}$ which are not very small compared to $\sigma$.
We have also given the algebraic expressions for magnetic fields at
relativistic limit $v=1$. 

The space-time profiles of electromagnetic fields in non-central Au+Au
collisions have been calculated based on the above analytic formula
as well as the exact numerical method. The UrQMD model was used to
simulate the space-time and momentum configurations of charged particles.
In collisions, the participant nucleons are treated differently from
spectators by introducing a smooth rapidity distribution to account
for newly produced charged particles in the central rapidity region.
The magnitudes of the axial components of both electric field and
magnetic field have the symmetry of flipping the signs of their transverse
coordinate arguments $x$ and $y$. But the magnitudes of transverse
components are only symmetric for flipping the sign of $x$ (in the
reaction plane) but not for $y$. This is the result of the CME. 

Combining the space-time evolution of electromagnetic fields with
hydrodynamic models or transport models, one can calculate in the
future the correlations of charged particles as possible observables
of the CME and compare with experimental data. 

\textit{Acknowledgments.} The authors thank L.G. Pang for helpful
discussions. Especially QW thanks K. Tuchin for insightful discussions
in the Workshop on Chirality, Vorticity and Magnetic Field in Heavy
Ion Collisions at UCLA. The authors are supported in part by the Major
State Basic Research Development Program (MSBRD) in China under Grant
2015CB856902 and 2014CB845406 respectively and by the National Natural
Science Foundation of China (NSFC) under the Grant 11535012. 

\bibliographystyle{apsrev}
\addcontentsline{toc}{section}{\refname}\bibliography{ref}

\end{document}